\documentclass[aps,pre,reprint,groupedaddress]{revtex4-1}

\usepackage{graphicx}
\usepackage{amsmath}
\usepackage{amssymb}
\usepackage{bm}
\usepackage{cancel}
\usepackage[normalem]{ulem}
\usepackage{xfrac}
\usepackage[colorlinks=true,linkcolor=blue,citecolor=blue,urlcolor=blue]{hyperref}

\usepackage{textcomp}
\usepackage{color}
\usepackage{xcolor}

\usepackage{hyperref}
\usepackage{cleveref}

\usepackage{subfigure}
\usepackage{placeins}
\usepackage{siunitx}
\usepackage{blindtext}

\newcommand{\miguel}[1] {\textcolor{blue}{#1}}

\begin{document}

\title{Emergent dynamics in 
 excitable flow systems}

\author{Miguel Ruiz-Garc\'ia}

\affiliation{Department of Physics and Astronomy, University of Pennsylvania, Philadelphia, PA 19104, USA}

\author{Eleni Katifori}

\affiliation{Department of Physics and Astronomy, University of Pennsylvania, Philadelphia, PA 19104, USA}

\date{\today}

\begin{abstract}
Flow networks can describe many natural and artificial systems. We present a model for a flow system that allows for volume accumulation, includes conduits with a non-linear relation between current and pressure difference, and can be applied to networks of arbitrary topology.
The model displays complex dynamics, including self-sustained oscillations in the absence of any dynamics in the inputs and outputs. 
In this work we analytically show the origin of self-sustained oscillations for the $1$D case. 
We numerically study the behavior of systems of arbitrary topology under different conditions: we discuss their excitability, the effect of different boundary conditions and wave propagation when the network has regions of conduits with linear conductance.
\end{abstract}
\maketitle

\section{Introduction}

Flow networks appear in a multitude of natural and artificial systems that require efficient distribution of nutrients, goods, or any other quantity of interest through the system. They are composed of a set of connections (e.g. resistors in the context of an electrical current network) that carry the flow between nodes. They also possess an external input (akin to an electric battery or a fluid pump) that provides the necessary energy for the flow to overcome energy losses due to dissipation. A widely studied example of a natural flow network is the vascular system of plants and animals whereas an artificial one is the power grid or water distribution systems. 

Since the seminal work of Kirchhoff in 1847 \cite{kirchhoff1847ueber}, much success has been achieved modelling flow systems as networks of linear resistors (see e.g. \cite{Murray1926}). These models present interesting physics, however, their linear nature, and absence of elements such as capacitors and inductors, assumes that a change on the boundary conditions (the net currents or pressures specified at a predetermined set of nodes, termed the \textit{contacts}) is instantaneously transmitted to the entire system. Thus, when the boundary condition (e.g. the voltage drop in the battery) is specified, there is a unique solution (up to a gauge) for the pressures throughout the system, and the dynamics of the whole system can be straighforwadly inferred from the dynamics imposed at the boundary nodes.

In this work we revisit the physics of resistor networks by relaxing the linearity condition. We consider a system of non-linear resistors, with a non-linear relation between the current flowing through a link $(i,j)$ and the pressure difference between the two nodes $i$ and $j$. We demonstrate that such a system can exhibit complex dynamics even in the absence of a time dependent drive. In our model, in contrast to simple linear resistor networks, the spatiotemporal variations of the current require that we consider transient internal storage of fluid. In the vascular system, for example, this is something that is accomplished by the dilation of the vessels. In our model we introduce this property by allowing, and accounting for, accumulation of volume at the network nodes. This enables 
the propagation of pressure and volume perturbations along the system.

The $1$D limit of our model is related to previous models used to study semiconductor superlattices~\cite{bonilla2005non,bonilla2010nonlinear}. These models can present complex dynamics such as self-sustained oscillations~\cite{bonilla2005non} or chaotic behavior~\cite{ruiz-garcia2017enhancing,essen2018parameter}. Also, its continuum limit has been used to study the Gunn effect in semiconductors~\cite{bonilla1995onset}.
Our model uses equations consistent with the scaling laws of fluidic systems rather than with semiconductor electronics. Unlike the previous work, it can be used on networks of arbitrary topology.

We focus our attention on the parameter range that displays oscillatory behavior under constant boundary conditions.  Moreover, the model presented in this work displays the basic properties characterizing excitable dynamical systems. In particular, for some parameter range, the system presents a stable point; and when perturbed away from it, it makes a large excursion in phase space before returning to the stable point. This is the signature behavior of excitable models~\cite{cross1993pattern,bonilla2010nonlinear}; we show an example of this behavior in section \ref{sec_exc} and in figure \ref{fig_excitability}. Other models of excitable networks have been previously studied, but these models usually include explicitly excitable elements at the nodes of the network. These elements can belong to different classes: they can be discrete variables that can be in a resting, excited or refractory state and that can excite their neighbors (see~\cite{kinouchi2006optimal}); or they can be neuron-like continuous variables whose dynamics are coupled to their neighbors' dynamics (see for example~\cite{roxyn2004selfsustained}). In our case the nodes are not intrinsically excitable, but just store volume. Excitability emerges as a global effect that stems from the combination of: (i) the coupling between the node capacity and the pressure field and (ii) the nonlinear conductance. This combination gives rise to the complex dynamics shown, at least in part, in this work.

The paper is structured in the following way. We first describe the mathematical model in section \ref{sec_math}. In section \ref{sec_3} we present the basic rationale for the emergence of spontaneous dynamics with an analytic study of the $1$D version of the model. Section \ref{sec_numerical_results} shows how to numerically solve the model (\ref{sec_num_int}) and different examples of interesting phenomenology (\ref{sec_robutness}). The discussion of our results is contained in section \ref{sec_discussion}. 

\begin{figure}
\includegraphics[width= \linewidth]{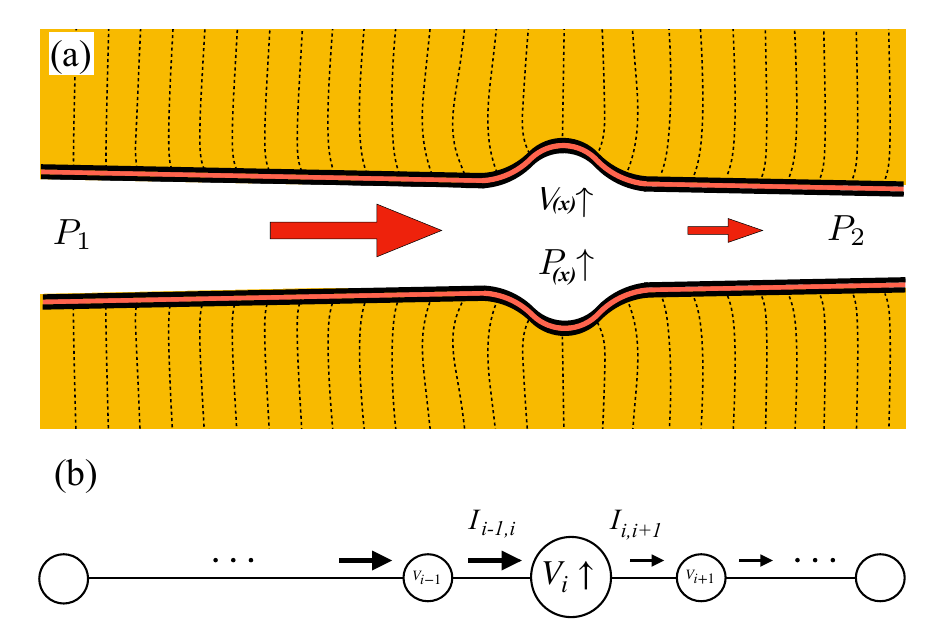}
   \caption{Sketch representing a section of a flexible tube (vessel) immersed in an elastic surrounding medium (a) and its network counterpart (b). (a) The pressure is set at the entrance and exit of the tube, driving flow from left to right (red arrows). The pressure in the tube decays from $P_1$ to $P_2$. Due to a non-linearity in the flow-pressure relation, present in biological systems (see Appendix B), or in microfluidic networks \cite{alvarado2017nonlinear,christensen2020viscous,louf2020bending}, the flow can diminish in one region of the tube (smaller arrow). This causes a volume accumulation in the preceding region at \textit{x} that subsequently makes the pressure to locally increase in the same region. This accumulation deforms the surrounding medium modifying the pressure also far from the accumulation region (see Appendix \ref{sec_coupling_pres_vol}). Note that, depending on the relation of the flow to the pressure gradient, this accumulation can decrease and vanish, returning the system to the original configuration, or it can grow to a new stable solution, such as a traveling wave. The dashed lines display the internal displacements in the surrounding medium due to the volume accumulation in the vessel. (b) The 1D discrete network modeling the continuous system in (a) conserves the basic ingredients that explain the emerge of the complex dynamical behavior. The volume $V_i$ is stored in the nodes $\{ i \}$ and it can vary with time, and the edges carry the currents $I_{i-1,i}$ between them.}
   \label{fig_sketch}
\end{figure}


\section{Model}
\label{sec_math}

\begin{figure}
\includegraphics[width=\linewidth]{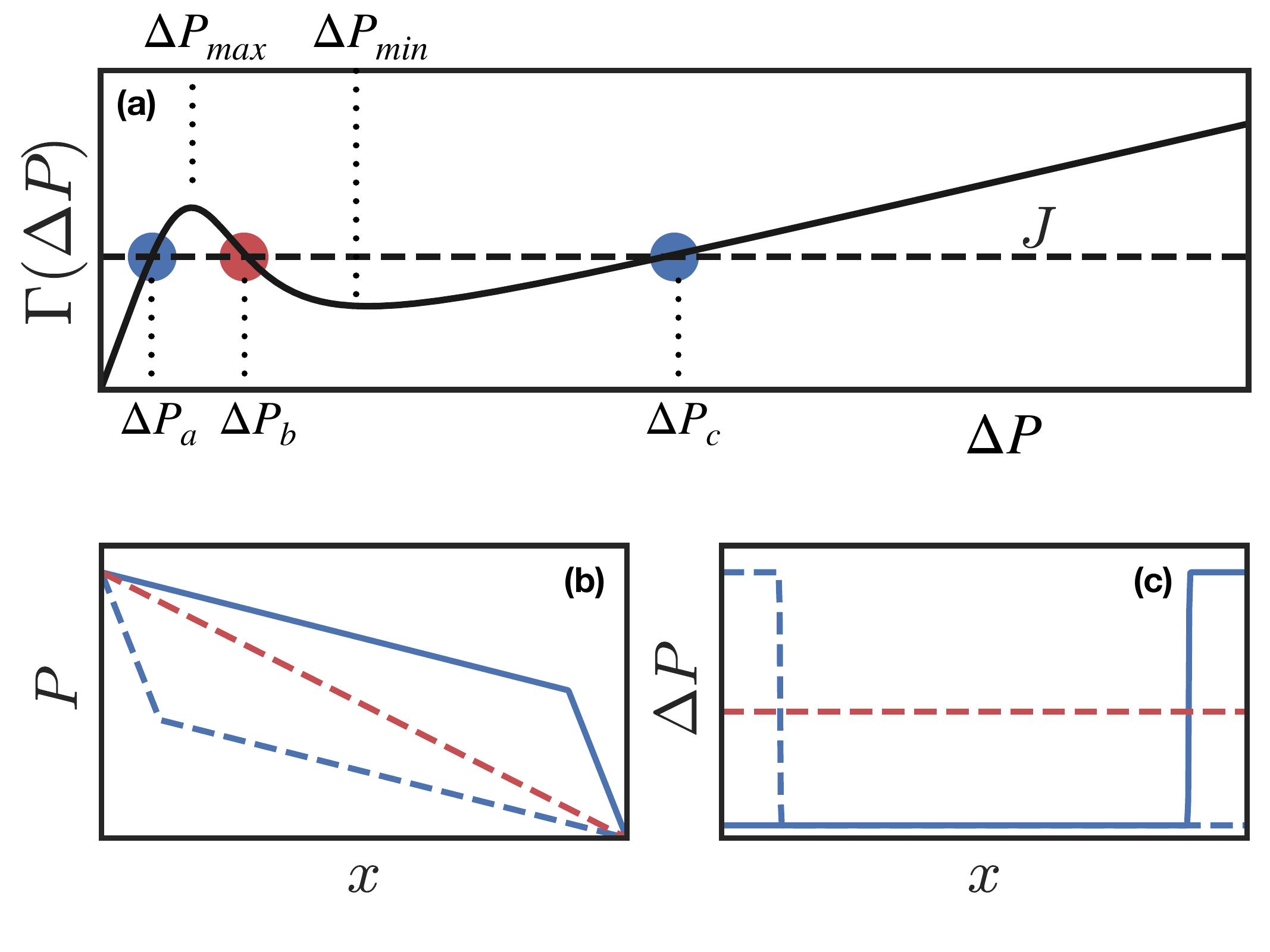}
\caption{Different plots depicting the construction of piece-wise profiles in the 1D model. (a) Generic plot of $\Gamma(\Delta P)$ versus $\Delta P$ (continuous line). The three highlighted points correspond to pressure differences $\Delta P$ such that $\Gamma(\Delta P_a)=\Gamma(\Delta P_b)=\Gamma(\Delta P_c)= J$. The red point ($\Delta P_b$) corresponds to an unstable configuration, whereas the two blue points ($\Delta P_a$ and $\Delta P_c$) are locally stable. Panels (b) and (c) show two piece-wise profiles in blue with a low pressure drop region ($\Delta P_a$) and a high pressure drop region ($\Delta P_c$). (b) and (c) also depict an unstable configuration with $\Delta P_i = \Delta P_b$, in dashed red lines. (b) $P$ versus $x$ (position on the $1$D network). The region of the curve with the shallow slope corresponds to $\Delta P_a$ and the steeper slope to $\Delta P_c$. (c) $\Delta P$ versus $x$. Note that in the $1$D case we use $\Delta P = P_i -P_{i+1}$, making $\Delta P$ positive for these profiles. The solid line in (b) and (c) corresponds to the case $\Delta P=\Delta P_a$ followed by $\Delta P=\Delta P_c$, and the dashed line to $\Delta P=\Delta P_c$ followed by $\Delta P=\Delta P_a$.}
\label{fig_sketches}
\end{figure}

The network model is composed of a set of nodes and connections (edges) between them (see Fig. \ref{fig_sketch} (b)). In contrast to the usual network of linear resistors, we allow for temporal accumulation and depletion of volume in the system. In a system of flexible tubes, the extra volume accumulated by a mismatch between the incoming and outgoing currents in a region of the system produces a temporary expansion of the tubes within this region, increasing their internal volume and compressing the surrounding medium, see Fig. \ref{fig_sketch} for a sketch showing this effect in a single tube. For simplicity, we account for these volume changes on the nodes of our network. Also, instead of a linear, Ohmic relationship between the pressure drop at the nodes and the current going through the edge that connects them, we consider the connections between the nodes as non-linear resistors with a region of negative slope (e.g. see figure \ref{fig_sketches}).

We define $V_i$ and $P_i$ as the volume and pressure at node $i$ of the network. We assume the following expression for the current $I_{ij}$ that goes from node $i$ to node $j$, through the edge that connects both nodes,
\begin{equation}
I_{ij} = 
    \begin{cases}
    V_i^2  \Gamma(\Delta P_{ij}) , \quad \text{if } P_i>P_j\\
     V_j^2  \Gamma(\Delta P_{ij}) , \quad \text{if } P_j>P_i
    \end{cases}
    \label{eq1}
\end{equation}
where the pressure drop is defined as \miguel{$\Delta P_{ij} = P_i - P_j$} and $\Gamma(\Delta P)$ is a general function of the pressure drop. 
For the sake of simplicity, unless otherwise noted, we only consider two kinds of vessels. The first is nonlinear on the pressure drop with one local maximum and one local minimum:
\begin{equation}
    \Gamma_{\mathrm{NL}}(\Delta P) =\gamma \frac{\Delta P_o^4+\epsilon \Delta P^4}{\Delta P_o^4+\Delta P^4} \Delta P,
    \label{eq2}
\end{equation}
where $\Delta P_o$ is a constant with units of pressure, see figure \ref{fig_sketches}. This phenomenological relation may play an important role in biological systems, see appendix \ref{sec_brain}; moreover, this type of non-monotonic relation can be also engineered in microfluidics devices, see for example \cite{christensen2020viscous,alvarado2017nonlinear}. In this work we will also use a linear relation on the pressure drop, 
\begin{equation}
    \Gamma_{\mathrm{L}}(\Delta P) = h \Delta P.
    \label{eq3}
\end{equation}
The parameter $\epsilon$ is a non-dimensional constant and $\gamma$ and $h$ are constants with dimensions  $([V][P][t])^{-1}$, where $[V],\ [P]$ and $[t]$ are the dimensions for volume, pressure and time, respectively. 
From equations \eqref{eq1}, \eqref{eq2} and \eqref{eq3} it follows that $I_{ij}$ is positive if $P_i>P_j$ and the current travels from $i$ to $j$, whereas it is negative if $P_i<P_j$ and the current travels from $j$ to $i$.

Note that we include a quadratic volume term in the expression for $I_{ij}$ and that $\Gamma_{\mathrm{NL}}$ is linear for low pressure drops, ensuring that we recover the scaling of Poiseuille flow at low pressures. In particular, considering the node volume $V_i$ in the network as a proxy for the volume stored in the region surrounding node $i$ in the real system, we have $V_i \propto R^2$. At sufficiently low $\Delta P_{ij}$, $ \Gamma(\Delta P_{ij})$ is linear in the pressure drop and \eqref{eq1} takes the form $I \propto  R^4 \Delta P$. Assuming that the length of the vessel ($l$) and the viscosity of the fluid ($\mu$) do not change, this scales as Poiseuille flow ($I = \pi R^4 \Delta P/8 \mu l$). Of course, depending on the specific characteristics of our system, this may be a strong approximation. Nevertheless, a wide range of exponents in the volume factor in equation \eqref{eq1} produce self-sustained oscillations, see supplementary materials. We also include a relation that couples volume and pressure,
\begin{equation}
    V_i -V_R = \alpha_d \sum_k L_{ik} P_k,
    \label{eq_vol_pres}
\end{equation}
where $V_R$ is a rest volume and $L_{ij}$ stands for the graph Laplacian.  $L_{ij}$ is equal to the degree of $i$ if $i=j$ and $-1$ if $i\ne j$ but $i$ and $j$ are connected by a link, $L_{ij}=0$ otherwise. This relationship, connecting pressure and volume, is a phenomenological expression consistent
with the physics of flow through an elastic medium (see Appendix A) and it is independent of the currents. When there is an accumulation of volume at one node, expression \eqref{eq_vol_pres} will result in a pressure distribution in the network  that produces currents that will promote the dispersion of the accummulation. When the volume decreases in one node with respect to its neighbors, the pressure field will promote currents that will increase the volume at that node. 

Finally, conservation of volume is imposed through:
\begin{equation}
    \frac{d V_i}{dt} = \sum_{k} \miguel{-} I_{ik},
    \label{eq_vol_deriv}
\end{equation}
where an increment of the volume at one node causes the drop of volume at neighboring nodes.

Without loss of generality we can make the equations dimensionless using $\tilde{V}\equiv \frac{V}{V_R}$, $\tilde{P}\equiv \frac{P}{\Delta P_o}$ and $\tilde{I} \equiv \frac{I}{2 \gamma V_R^2 \Delta P_o}$, which sets the dimensionless time to be $\tilde{t} \equiv 2V_R \gamma \Delta P_o t$. With these substitutions,  equations ~\cref{eq1,eq2,eq3,eq_vol_pres,eq_vol_deriv} become:

\begin{equation}
\tilde{I}_{ij} = 
    \begin{cases}
    \frac{1}{2} \tilde{V_i}^2  \tilde{\Gamma}(\Delta \tilde{P}_{ij}) , \quad \text{if }  \tilde{P}_i> \tilde{P}_j\\
     \frac{1}{2} \tilde{V_j}^2 \tilde{\Gamma}(\Delta \tilde{P}_{ij}) , \quad \text{if }  \tilde{P}_j> \tilde{P}_i
    \end{cases}
    \label{eq_currents}
\end{equation}

\begin{equation}
    \tilde{\Gamma}_{\mathrm{NL}}(\Delta \tilde{P}) = \frac{1+\epsilon \Delta \tilde{P}^4}{1+\Delta \tilde{P}^4} \Delta \tilde{P},
    \label{G_nl}
\end{equation}

\begin{equation}
    \tilde{\Gamma}_{\mathrm{L}}(\Delta \tilde{P}) = \tilde{h} \Delta \tilde{P},
\end{equation}

\begin{equation}
    \tilde{V}_i -1 = \frac{\alpha_d \Delta P_o}{V_R} \sum_k L_{ik} \tilde{P}_k,
    \label{eq_dim_vp}
\end{equation}

\begin{equation}
    \frac{d \tilde{V}_i}{d\tilde{t}}= -\sum_{k} \tilde{I}_{ik},
    \label{eq_dl_dv}
\end{equation}

We will also define $\alpha \equiv \frac{\alpha_d \Delta P_o}{V_R}$ so that \eqref{eq_dim_vp} takes the form:
\begin{equation}
    \tilde{V}_i -1 = \alpha \sum_k L_{ik} \tilde{P}_k.
    \label{eq_dl_V}
\end{equation}
In the rest of this work we will drop the tildes for the sake of clarity. Further considerations on alternatives expressions for $\Gamma(\Delta P)$ or the possible variations of equation \eqref{eq_vol_pres} may increase the complex dynamical behavior of the model.


\section{Analytical results: Stability and wave propagation in 1D}
\label{sec_3}

In this section we use the $1$D version of our model to explain some of the dynamics exhibited by this system. Here, to simplify the formulae we redefine the pressure difference between two nodes as $\Delta P_i = P_i - P_{i+1}$, and we adopt this sign convection for the rest of this section. Some of the arguments presented here are inspired by the analytical work on semiconductor superlattices. Useful review references of that work can be found in \cite{bonilla2005non,bonilla2010nonlinear}.

\subsection{Stability of homogeneous stationary profiles }

\label{sec_homogeneous}

If all the edges (non-linear resistors) are equivalent, and there is a constant pressure drop at every edge, then $\Delta P_i = \Delta P^*$ and $V_i=1$. This results in a constant current across the system, a stationary point of the dynamics. For simplicity lets consider here a generic expression for the current $I(\Delta P_i)$ from node $i$ to $i+1$ that only depends on the pressure difference between the two nodes. We also use the coupling between pressure and volume, equation \eqref{eq_dl_V}, which in the $1$D network takes the form,
\begin{equation}
    V_i -1 = \alpha (\Delta P_i - \Delta P_{i-1}).
    \label{eq_DV_a}
\end{equation}
Consider now a small perturbation around the stationary state,
\begin{equation}
    \Delta P_i = \Delta P^* + \epsilon \Delta p_i\text{, and } V_i = 1 + \epsilon v_i.
    \label{eq_pert}
\end{equation}
Substituting these expressions into Eq. \eqref{eq_DV_a} we get,
\begin{equation}
    v_i=\alpha(\Delta p_i-\Delta p_{i-1}).
    \label{eq_pert2}
\end{equation}
The conservation of volume in the system is given by
\begin{equation}
    \frac{d V_i}{dt} = I(\Delta P_{i-1})-I(\Delta P_{i}).
\end{equation}
Linearizing $I$ around $\Delta P^*$ and using \eqref{eq_pert} and \eqref{eq_pert2} we get
\begin{equation}
    \frac{dv_i}{dt} = - \frac{I'(\Delta P^*)}{\alpha} v_i.
\end{equation}
Now it is clear that a negative slope of $I(\Delta P)$ at $\Delta P^*$ will result in an exponential increase of the small perturbations of the accumulated volume ($v_i$). This is the basic mechanism that renders some of the ``trivial'' stationary solutions of the model unstable when the current versus pressure drop  presents a region of negative slope.

\subsection{Piece-wise linear profiles}
\label{sec_piece_wise}

We now explore when piece-wise constant pressure drop profiles are stationary solutions of the dynamics.
We consider again a system of $N$ edges arranged on a line. A stationary solution requires the current from one node to the next one to be constant throughout the whole system. As we have seen in the previous section, if the boundary conditions are the constant external pressures $P_0 = \Pi = N \Delta P_b$ and $P_{N}=0$, such that $\Delta P_b$ lays on the negative-slope region of $\Gamma (\Delta P)$, the solution $\Delta P_i=\Delta P_b,  \forall i \in (0,N)$ is unstable (red dot on Fig. \ref{fig_sketches} (a)). However, if $\Gamma (\Delta P)$ presents a local maximum followed by a local minimum, as in Fig. \ref{fig_sketches} (a), we can build a different pressure profile containing two regions of constant pressure drop $\Delta P_a$ and $\Delta P_c$ (blue points in Fig. \ref{fig_sketches} (a)). These piece-wise pressure profiles are presented in Fig. \ref{fig_sketches} (b) and (c) as continuous and dashed blue lines. To study if these profiles are stationary solutions of our model, let us assume first that the transition between the regions of low ($\Delta P_a$) and high ($\Delta P_c$) pressure drop happens at a single point. This point is situated at $k$:
\begin{equation}
    k=\frac{\Pi - N\Delta P_c}{\Delta P_a-\Delta P_c},
\end{equation}
for the continuous line in figures \ref{fig_sketches} (b) and (c), and at $N-k$ for the blue dashed line. Within the regions with constant pressure drop ($\Delta P_a$ and $\Delta P_c$), the volume is $1$ and the current is equal to $J$ (see figure \ref{fig_sketches} (a)). However, exactly at the node where the slope of the pressure field changes from $\Delta P_a$ to $\Delta P_c$, the volume is different from $1$ (due to equation \eqref{eq_DV_a}). According to Eq. \eqref{eq_currents}, $I_{k,k+1} \ne J$ while $I_{k-1,k} = J$, and as result $\frac{d V_k}{d t} \ne 0$, showing that this configuration is not a stationary solution of the dynamics.

We will now examine a configuration where the transition between the regions with pressure drop $\Delta P_a$ and $\Delta P_c$ spans multiple nodes. Like in the previously examined case, we know that the volume at the node situated in the right extreme of the transition region should be $1$, to avoid a mismatch in the currents. This can only be achieved exactly in the continuous limit, however let us consider the case where the transition region spans a finite number of nodes.
 
Again, a stationary solution should present a constant current throughout the entire system. Within the constant pressure drop regions $V_i=1$, and the current is then $J$. A stationary solution requires the current in the intermediate region to be,
\begin{equation}
    V_{i}^2 \Gamma (\Delta P_{i}) = J,
    \label{eq_a29}
\end{equation}
where
\begin{equation}
    V_i = 1 + \alpha (\Delta P_i - \Delta P_{i-1}).
    \label{eq_a27}
\end{equation}
We can rewrite equation \eqref{eq_a29} as, 
\begin{align}
    [1+\alpha^2 (\Delta P_i -\Delta P_{i-1})^2 + 2 \alpha (\Delta P_i -\Delta P_{i-1}) ] \Gamma (\Delta P_i) = J.
    \label{eq_jjj}
\end{align}
Defining 
\begin{equation}
  \theta_i = \alpha (\Delta P_i -\Delta P_{i-1}),
  \label{eq_theta_1}
\end{equation}
equation \eqref{eq_jjj} takes the form,
\begin{equation}
    \Gamma (\Delta P_i) \theta_i^2 +  2\Gamma (\Delta P_i) \theta_i +(\Gamma (\Delta P_i)-J) = 0,
\end{equation}
which has the solution,
\begin{equation}
    \theta_i = -1 + \sqrt{J/\Gamma (\Delta P_i)}.
    \label{eq_theta}
\end{equation}
Note that we only keep the positive sign of the square root since  $1+\theta_i$ is the volume at node $i$.
Using \eqref{eq_theta_1} we can write,
\begin{equation}
    \Delta P_i = \Delta P_{i-1} + \frac{1}{\alpha}( -1 + \sqrt{J/\Gamma (\Delta P_i)}).
    \label{eq_ps}
\end{equation}
Let us now consider two cases, a piece-wise pressure profile that connects a region of $\Delta P_a$ to a region of $\Delta P_c$ and the reverse (going from $\Delta P_c$ to $\Delta P_a$). In the former case, the transition region is characterized by a $\Delta P_i = P_i - P_{i+1}$ that grows as $i$ increases. If $\Gamma (\Delta P_i)$ has a local maximum at $\Delta P_{max}$ followed by a local minimum at $\Delta P_{min}$, as displayed in figure \ref{fig_sketches},  equation \eqref{eq_ps} will have a solution for $\Delta P_i$ that grows from $\Delta P_a$ to $\Delta P_c$ only if 
\begin{equation}
    \alpha \leq \frac{-1+\sqrt{J/\Gamma (\Delta P_{min})}}{\Delta P_{min} - \Delta P_a} \equiv \alpha_{c1},
    \label{eq_alpha_1}
\end{equation}
(see supplementary materials for the derivation).
Similarly, in the case where a piece-wise pressure profile connects a region of $\Delta P_c$ to a region of $\Delta P_a$, $\Delta P_i = P_i - P_{i+1}$ must decrease in the transition region as $i$ increases. Following similar arguments as before, \eqref{eq_ps} will have a solution if 
\begin{equation}
    \alpha \leq \frac{1-\sqrt{J/\Gamma (\Delta P_{max})}}{\Delta P_{c} - \Delta P_{max}}\equiv \alpha_{c2}.
    \label{eq_alpha_2}
\end{equation}
If $\alpha$ does not satisfy \eqref{eq_alpha_1} or \eqref{eq_alpha_2} these piece-wise pressure profiles cannot be stationary solutions of the system. 

In summary, when a constant pressure drop $\Pi$ is applied to the network such that $\Pi/N=\Delta P_b$, with $\Delta P_b$ in the negative-slope region of $\Gamma (\Delta P)$, the homogeneous pressure profile ($\Delta P_i=\Delta P_b, \forall i$) is unstable. In addition, if $\alpha > \alpha_{c1}$ and $\alpha > \alpha_{c2}$, the piece-wise pressure profiles discussed above are not stationary solutions. Performing numerical simulations in $1$D networks (see supplementary materials) we observe self-sustained oscillations outside these regions of stationary solutions. Next section provides a qualitative explanation of their behavior. 

\subsection{Travelling piece-wise profiles}

\label{sec_travelling_piec}


\begin{figure*}
    \centering
    \includegraphics[width=\textwidth]{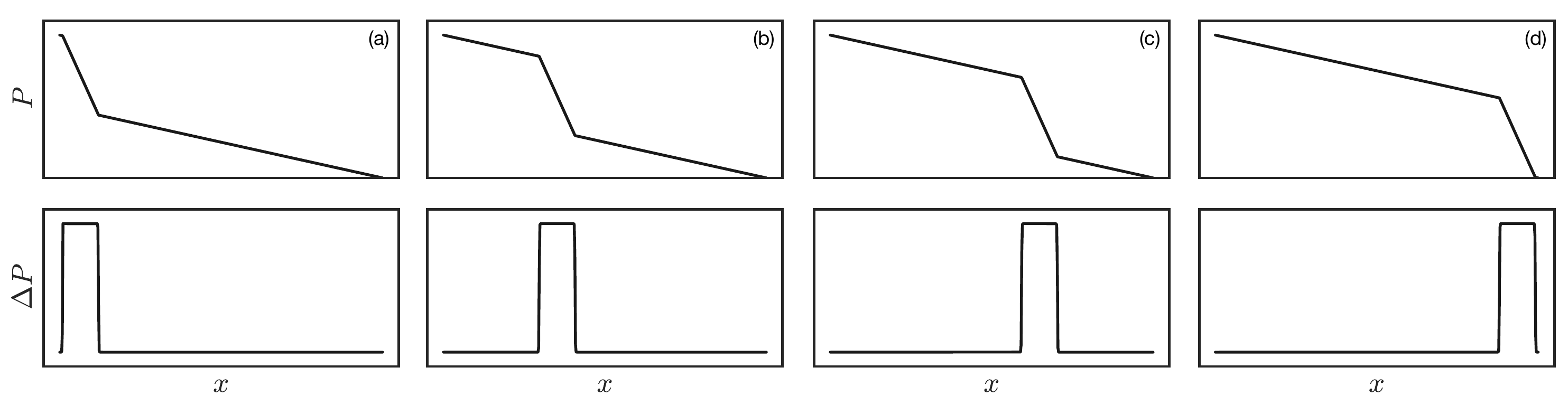}
    \caption{A travelling wave through a system with constant pressure boundary conditions. The upper half of every panel depicts the pressure at every node of the system whereas the bottom half shows the pressure drop at each node.
    These pressure profiles contain two transitions between $\Delta P_a$ and $\Delta P_c$. The four panels correspond to different snapshots. The wave travels from left to right maintaining its shape (from panel (a) to (d)). The wave can recycle once it reaches the end of the domain, satisfying the boundary conditions (a constant external pressure drop $\Pi$) at all times.}
    \label{fig_trav}
\end{figure*}

We have shown that the piece-wise profiles shown in figure \ref{fig_sketches} (b) and (c) can only be stationary solutions of our model for $\alpha$ smaller than certain values. However, we can wonder if other piece-wise profiles could move with some velocity across the system as travelling waves. An instance of this behavior is displayed in figure \ref{fig_trav} where there is a pressure profile that travels from left to right while satisfying the boundary condition, a constant external pressure difference between the first and last node.

Consider again a piece-wise profile where $\Delta P_i$ changes within an intermediate region from $\Delta P_a$ to $\Delta P_c$. Since in that intermediate region $\Delta P_{i-1} < \Delta P_{i} $ we know $V_i>1$ and this is a region of volume accumulation. If this volume accumulation were to be rigidly translated with speed $c_a$ across the system, we can expect the pressure drop at any node within the system to change with time as $\Delta \dot{ P}_i \sim -c_a (\Delta P_i -\Delta P_{i-1})$. Using equations \eqref{eq_vol_deriv} and \eqref{eq_a27}, one can define $I$, which is constant throughout the system,
\begin{equation}
    I := \alpha \Delta \dot{P}_i +V_{i}^2 \Gamma (\Delta P_{i}) = \alpha \Delta \dot{P}_{i-1} +V_{i-1}^2 \Gamma (\Delta P_{i-1}).
    \label{eq_J}
\end{equation}
Using \eqref{eq_J},
the difference in pressure drop ($\Delta P$) between the beginning and the end of the intermediate region can be rewritten as,
\begin{align}
    &\Delta P_c - \Delta P_a = \sum_i  (\Delta P_i - \Delta P_{i-1})=  \sum_i  \frac{-1}{c_a} \Delta \dot{P}_i =   \nonumber\\
    &=\sum_i  \frac{-1}{c_a\alpha} [I-V_i^2 \Gamma(\Delta P_i)] = \frac{1}{c_a\alpha}  \Big\{ \sum_i [\Gamma(\Delta P_i)-I] + \nonumber\\
    & + \sum_i [ \alpha^2 (\Delta P_i -\Delta P_{i-1})^2 + 2 \alpha (\Delta P_i -\Delta P_{i-1}) ]\Gamma(\Delta P_i) \Big\}.
\end{align}
Solving for the velocity ($c_a$) we get,
\begin{align}
    &c_a = \frac{1}{(\Delta P_c - \Delta P_a)\alpha}  \Big\{ \sum_i [\Gamma(\Delta P_i)-I] + \nonumber\\
    & + \sum_i [ \alpha^2 (\Delta P_i -\Delta P_{i-1})^2 + 2 \alpha (\Delta P_i -\Delta P_{i-1}) ]\Gamma(\Delta P_i) \Big\}.
 \label{eq_ca_1}
\end{align}
One can follow similar steps for a profile where $\Delta P_i$ goes from $\Delta P_c$ to $\Delta P_a$ within the intermediate region (a volume depletion region) obtaining a similar expression $c_d$.

In general $c_a$ and $c_d$ can be different. The specific shape of the pressure profile ($P_i$ and therefore $\Delta P_i$) within these intermediate regions control the value of $c_a$ and $c_d$. In fact, one accumulation and  one depletion region can adapt their shapes, and thus their velocities, to travel together with the same speed. This creates a soliton-like travelling wave that always satisfies a constant pressure drop between the beginning and end of the system. This is schematically shown in figure \ref{fig_trav}. More detailed asymptotic analysis, as the ones performed for semiconductor dynamics lay outside the scope of this work. Asymptotic analysis concerning electronic dynamics in semicondutor heterostructures, which are described with a related model to our $1$D case, can be found in~\cite{bonilla2010nonlinear}.


\section{Numerical results on networks of arbitrary topology}
\label{sec_numerical_results}

\subsection{ Boundary conditions and time integration of the model}
\label{sec_num_int}

The evolution of the system with time is determined by the evolution of the pressure field at every node $P_i$. Taking the derivative of equation \eqref{eq_dl_V} with respect to time and using equation \eqref{eq_dl_dv} we get,
\begin{equation}
    \alpha \sum_k L_{ik} \dot{P}_k = - \sum_k I_{ik}.
    \label{eq_p_i}
\end{equation}
To solve for $\dot{P_i}$ we need to consider the boundary conditions of the system of equations, i.e. we select $n$ nodes from the system, set them as the contact points, and externally control their pressure. We assume that these nodes are reservoirs with a constant volume ($V_n = 1$). We augment the graph Laplacian by including the pressure boundary conditions as new rows and columns. For example, for the case where we consider two pressure contacts at nodes $n_1$ and $n_2$ we have:
\begin{equation}
\hat{L}_{kl}= \begin{bmatrix}
  &  &  &   &  &  \vdots  &  \\
  & &   &   &  &  0  & \vdots \\
  &  &  L_{ij}  &  &   & 1 & 0 \\
   & &    &  &   &  0 & 1 \\
  &   &   &  &   &  \vdots & 0 \\
  &  \dots &  0 &1  & 0 & \dots &   \vdots \\
   &  &  \dots  &  0 & 1 & 0 &  \dots
\end{bmatrix},
\end{equation}
where the elements of the two new rows and columns are all zero except for $\hat{L}_{N+1,n_1} = \hat{L}_{N+2,n_2} = \hat{L}_{n_1,N+1} = \hat{L}_{n_2,N+2} = 1$. We also add corresponding elements to $P_i$ and to the vector of currents,
\begin{equation}
    \dot{\hat{P}}_k= \begin{bmatrix}
             \dot{P}_i    \\
         \lambda_1 \\
         \lambda_2
\end{bmatrix}, \quad    
\hat{Q}_i= \begin{bmatrix}
             -\sum_k I_{ik}    \\
         \beta_1 \\
         \beta_2
\end{bmatrix},
\label{eq_p_i_2}
\end{equation}
to finally get:
\begin{equation}
    \dot{\hat{P}}_j = \frac{1}{\alpha} \sum_i \hat{L}_{ji}^{-1}  \hat{Q}_{i}.
    \label{eq_dyn}
\end{equation}

Note that in this formulation, from equations \eqref{eq_p_i_2} and \eqref{eq_dyn}, $\lambda_i$ (with $i=1,2$) is related to the net currents going in (or out) of the system at node $n_i$, whereas $\beta_i/\alpha$ is the rate of change of the imposed pressure at the contact nodes. In general we will use as initial conditions $P_i=0$ and $V_i = 1 \ \forall i$. We carry out the time integration as follows.  Starting from $P_{n_1} = 0$, we increase the pressure of $n_1$ at constant rate ($\beta_1 = const$, $\beta_2 = 0$) until it reaches the desired value $\Pi$. Then we set $\beta_1 =\beta_2 =0$. Note that controlling $\beta_i$ as a function of time enables to freely control the pressure at the contacts ($n_1$ and $n_2$). Integrating the system of equations contained in expression \eqref{eq_dyn} with time we get the evolution of the pressure and volume in the system. Additionally, we obtain $\lambda_0$ and $\lambda_1$.

\begin{figure}
\includegraphics[width= \linewidth]{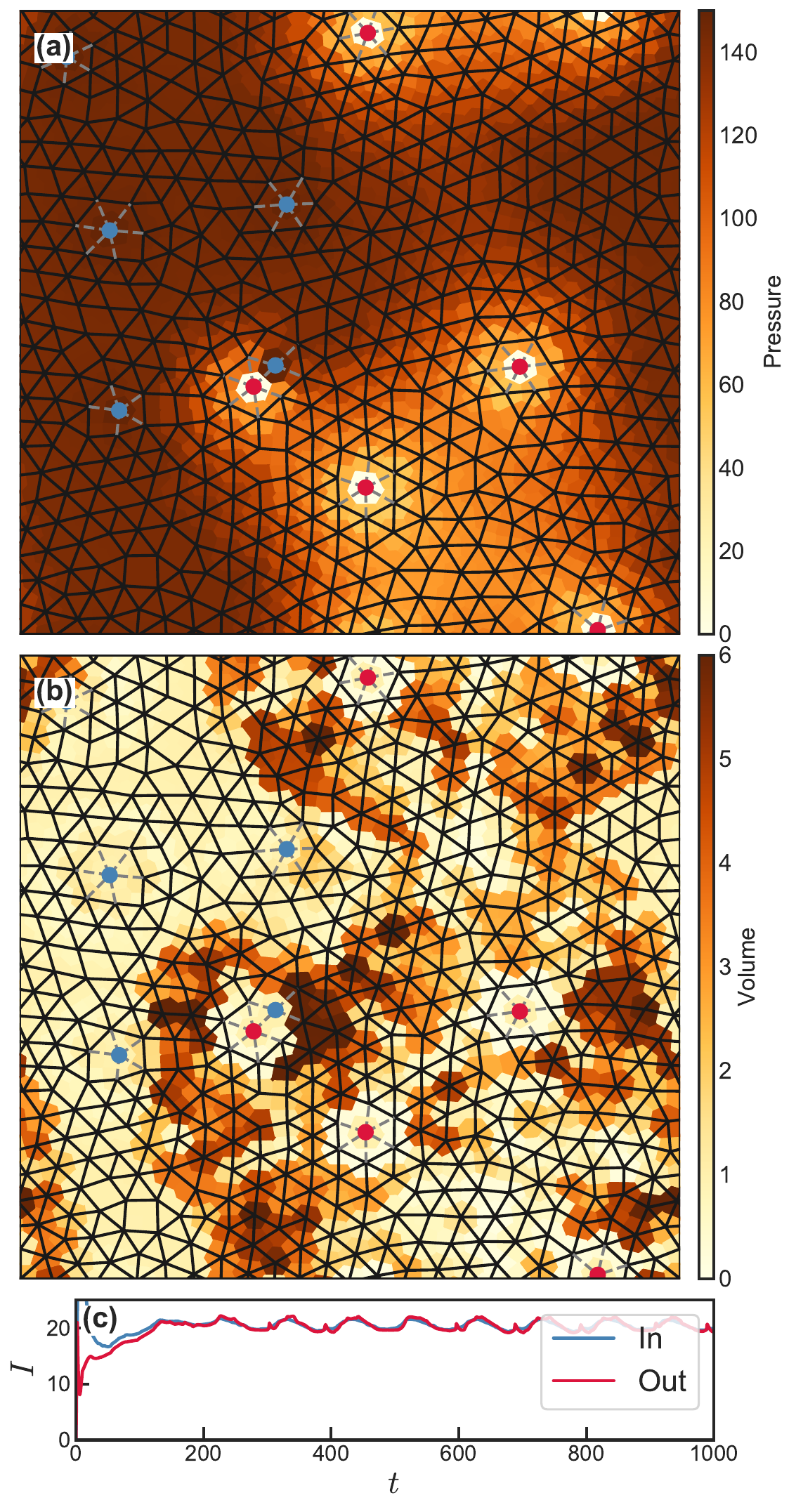}
   \caption{Volume waves on a disordered planar network. The two upper panels show the pressure (a) and volume (b) distribution at dimensionless time $1000$. The value of $P_i$ and $V_i$ is represented by the color surrounding each node. The network has $512$ nodes and an average degree of $\sim 5.5$. We have placed $10$ contacts at random (blue and red dots). The pressure increases at constant rate, $\beta_1=5$, at the blue contacts until they reach a value of pressure $P = 150$ and remains constant afterwards. The pressure at the red contacts is always $0$. All edges in the network follow $\Gamma_{NL}$ with $\epsilon = 0.001$ (continuous lines), except for the edges connected to one of the contacts that follow $\Gamma_L$ with $h=1/5$ (dashed lines). We used $\alpha = 0.32$. Panel (c) shows the evolution of the current that is going in (blue) and out (red) of the system with respect to time. Note that the system is undergoing self-sustained oscillations where volume waves travel across the system (with periodic boundary conditions) until they reach another contact.}
   \label{fig_time_evolution}
\end{figure}

For an example of time integration see Fig. \ref{fig_time_evolution}: (a) and (b) respectively show a snapshot of the pressure and volume distribution in a planar disordered network, whereas panel (c) shows the total current at the contacts that is going in and out of the system with time. For this simulation we chose $10$ pressure sources at random, in five of them (red nodes) the pressure is $0$ during the whole simulation. For the other five contacts (blue nodes) the pressure is ramped up at a fixed rate to a constant value of $150$ and kept constant afterwards. After a brief transient, the system exhibits stable disordered volume waves that travel through the system all while satisfying time independent pressure boundary conditions.  The panel (c) of figure \ref{fig_time_evolution} shows oscillations in the total current that is going in and out of the system.

\subsection{Results}
\label{sec_robutness}

The aim of this section is to present and explain  different instances of the rich behavior that this model can display in 1D and 2D. We begin by an exploration of the behavior of the 1D network, where we can directly compare our analytical predictions in Sec~\ref{sec_3} with the results of the simulation. For the 1D system we present a phase diagram that summarizes the different types of behavior the system can exhibit. We also demonstrate that the network can behave as an excitable medium. We then move to 2D networks, where we first demonstrate the highly complex patterns of dynamics that can be present when more than two contact points are present, a behavior that is intrinsically absent from 1D.  Last we demonstrate that in 2D the system exhibits qualitative similarities in behavior with excitable systems such as the one in ~\cite{Zykov2017}. 

\subsubsection{Phase diagram and robustness of the dynamical behavior}

\begin{figure}

\includegraphics[width=\linewidth]{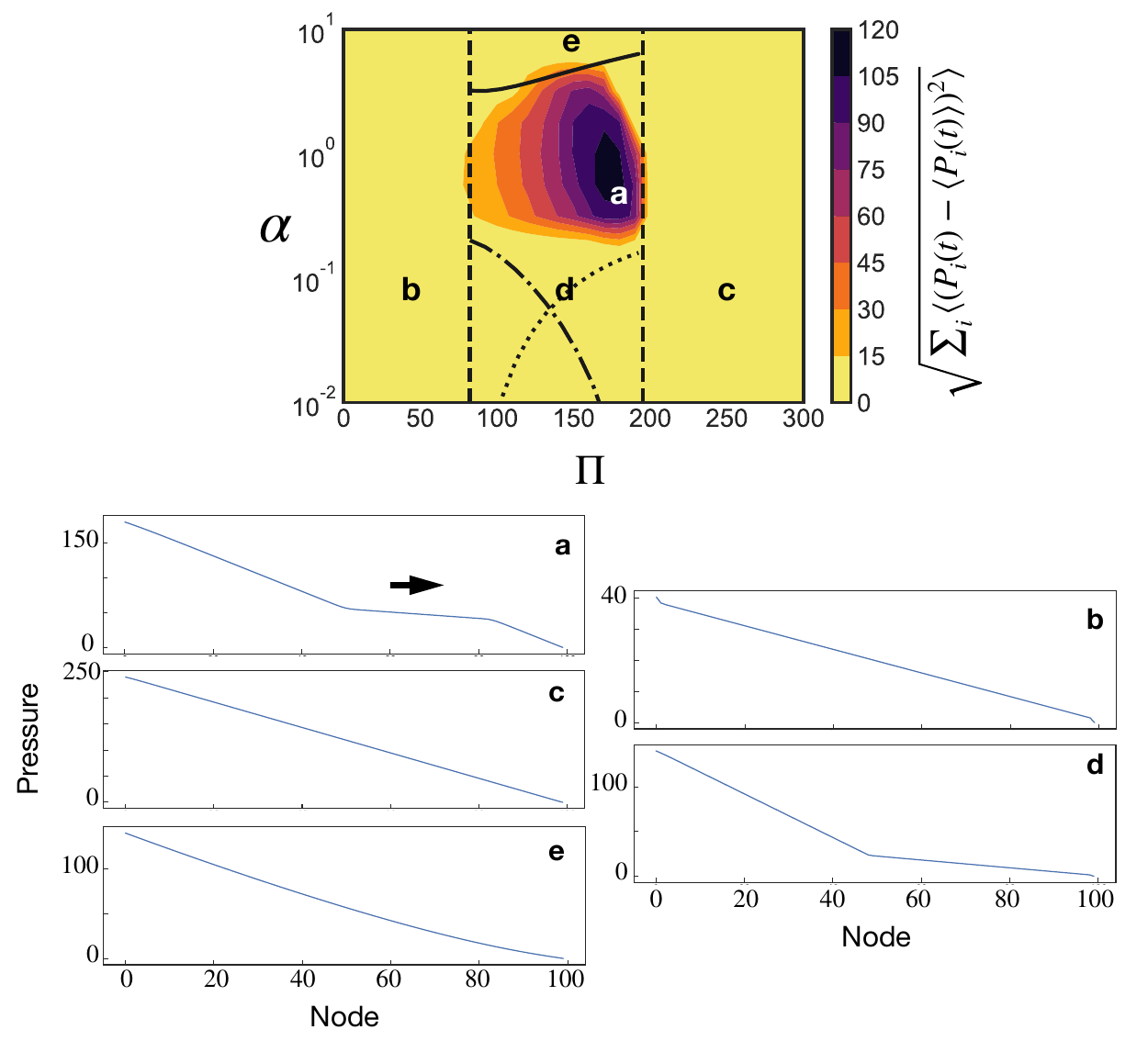}
\caption{Phase diagram for a $1$D network  of $100$ nodes, $\epsilon = 0.15$ and $h=1/5$. Colormap indicates the amplitude of the oscillations. Vertical dashed lines correspond to $100 \Delta P_{min}$ and  $100 \Delta P_{max}$. Dotted and dashed-dotted lines correspond to expressions \eqref{eq_alpha_1} and \eqref{eq_alpha_2}, respectively. The continuous line corresponds to expression \eqref{eq_ine_2}. Note how the oscillations occur in the region delimited by the analytical bounds. Panels below the phase diagram show the pressure profile throughout the network at the points marked with letters a-e. (a) shows a snapshot of the time-dependent pressure profile that corresponds to the region of largest oscillations. Note that it has the structure proposed in figure \ref{fig_trav}; (b) and (c) show two linear stationary pressure profiles, as the theory predicts; (d) depicts a piece-wise stationary profile, as described in section  \ref{sec_piece_wise}; finally, (e) shows a stationary pressure profile with curvature ($V_i<1$), as our analytical results predict for large values of $\alpha$.
}

\label{fig_phase_diag_1d}
\end{figure}

\begin{figure}
\includegraphics[width= 0.9\linewidth]{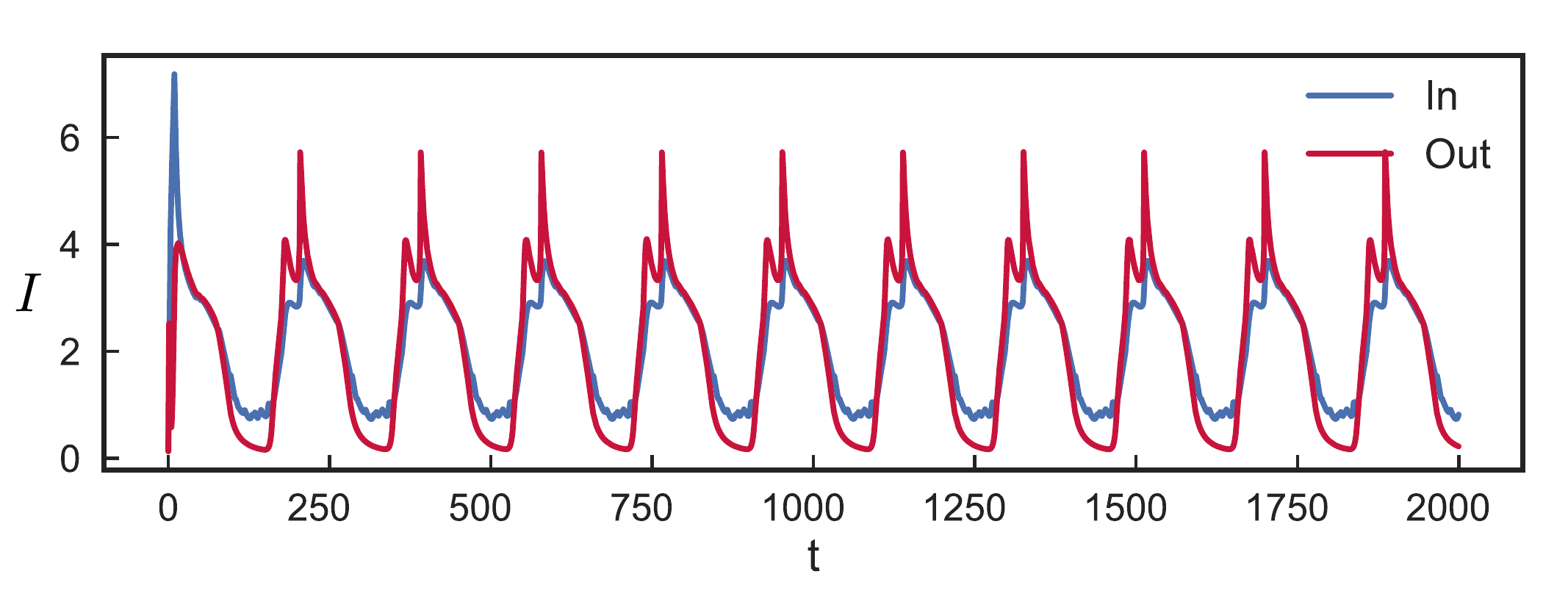}
\caption{Self-sustained oscillations in a regular cubic network of dimensions $7$x$7$x$7$ with two contacts connected to two opposite corners.
Plot shows the current going in and out of the system during a simulation displaying self-sustained oscillations. Parameter values are $\alpha = 0.32$, $\Pi = 150$, $\epsilon = 0.001$, $h=1/5$, $\beta_1=5$ ($\beta_1=0$ when $P_1$ reaches the desired value) and $\beta_2=0$.}
\label{fig5}
\end{figure}

In Fig. \ref{fig_phase_diag_1d} we present a phase diagram for a $1$D network of $N=100$ nodes. For each value of $\alpha$ and $\Pi$ we perform an independent simulation. $\Pi$ is the pressure difference between the two extremes, the  contact points. The initial conditions are $V_i=1 \text{ and} \ P_i=0,\ \forall i$. For $t>0$ we ramp up the pressure at one contact point until it reaches $\Pi$ and keep it fixed afterwards. The pressure at the other contact point is fixed at $0$. The protocol of gradual ramping up of the pressure at the contact point was chosen because of its connection to experiments (the system starts disconnected from the pressure source, so that all internal points are initially at zero pressure). The arbitrary choice of initial conditions generally only affects the transient and not the eventual dynamic or stationary steady state.

Figure \ref{fig_phase_diag_1d} exhibits a region of oscillatory behavior. The phase diagram also contains the analytical predictions that constrains the region where we expect to see oscillations, in good agreement with the numerical results. The two vertical dashed lines limit the region at which the homogeneous stationary solution is unstable, see section \ref{sec_homogeneous}. For larger or smaller pressure differences ($\Pi$) we expect to see linear pressure drops, as panels \ref{fig_phase_diag_1d} (b) and (c) confirm. Below the dotted and dashed-dotted lines piece-wise stationary profiles are stable, see section \ref{sec_piece_wise}. This is in good agreement with panel \ref{fig_phase_diag_1d} (d). Panel \ref{fig_phase_diag_1d} (a) shows a snapshot of the time evolution of the pressure profile for a point in the phase diagram where the system displays self-sustained oscillations. The pressure distribution in this case is formed by three linear pieces, as described in section \ref{sec_travelling_piec} (see also figure \ref{fig_trav}).

Finally, we would like to understand why the oscillations disappear for large $\alpha$. As shown in figure \ref{fig_trav} and \ref{fig_phase_diag_1d} (a), self-oscillations occur for piece-wise solutions that contain two transition regions, one where $V_i>1$ and another where $V_i<1$. 
For a total pressure decay $\Pi= N \Delta P_b$ (where $\Delta P_{min}<\Delta P_b<\Delta P_{max}$), we know that the pressure drop inside the accumulation region ($V_i>1$) goes from $\Delta P_a$ to $\Delta P_c$ whereas the pressure drop inside the depletion region ($V_i<1$) goes from $\Delta P_c$ to $\Delta P_a$, see figure \ref{fig_sketches}. Moreover, we know that volume has to be always positive. Using \eqref{eq_DV_a}, and $V_i>0$ we get
\begin{equation}
    1 +\alpha (\Delta P_i - \Delta P_{i-1}) > 0.
    \label{eq_ine_1}
\end{equation}
When the system displays oscillations, the transition regions should occupy a small portion of the network, see section \ref{sec_travelling_piec} and figure  \ref{fig_phase_diag_1d} (a). A traveling wave has two narrow depletion/accumulation regions separated by a linear pressure drop domain. Such a traveling wave cannot be maintained if the depletion/accumulation regions are comparable to the size of the system. If we suppose that the depletion region occupies approximately $10\%$ of the entire system, or $N/10$ nodes, we can approximate \eqref{eq_ine_1} by
\begin{equation}
    1 +\alpha \frac{\Delta P_a - \Delta P_c}{N/10} > 0,
    \label{eq_ine_2}
\end{equation}
what translates to
\begin{equation}
    \alpha  <  \frac{10}{\Delta P_c - \Delta P_a},
    \label{eq_ine_2}
\end{equation}
for $N=100$. We include expression \eqref{eq_ine_2} in figure \ref{fig_phase_diag_1d} with a continuous line, below which we expect to see oscillations. According to this, for $\alpha$ above the continuous line the depletion region ($V_i<1$) has to occupy a larger fraction of the nodes of the network, to be able to respect the $V_i>0$ condition. This does not allow the traveling wave to develop. This is in good agreement with figure \ref{fig_phase_diag_1d} (e) that shows a depletion region that occupies almost the complete network (note the subtle curvature of the pressure profile what implies $V_i<1$).

The maximum fraction of the traveling wave occupied by the accumulation and depletion regions was estimated at $10\%$. However, note that a factor of $2$ increase or decrease in that fraction would still provide qualitatively good agreement with the simulation, as the region of the phase diagram that exhibits oscillations spans alsmost two decades.

Additionally, our work indicates that the emergence of complex dynamics in this model is a robust effect that persists after modifying different properties of the system. To show this, we present in the supplementary materials phase diagrams for $1$D networks, with different shapes of $\Gamma_{NL}$ and $\Gamma_L$ and different distributions of linear edges. Self-sustained complex dynamics are found for a broad range of $\alpha$ and $\Pi$ values. Complex dynamics are also present in non-planar networks. To illustrate this, we include here a simulation carried out using a cubic lattice, also displaying self-sustained oscillations, see figure \ref{fig5}. 

\subsubsection{Excitability}
\label{sec_exc}

\begin{figure}
    \centering
\includegraphics[width=0.9 \linewidth]{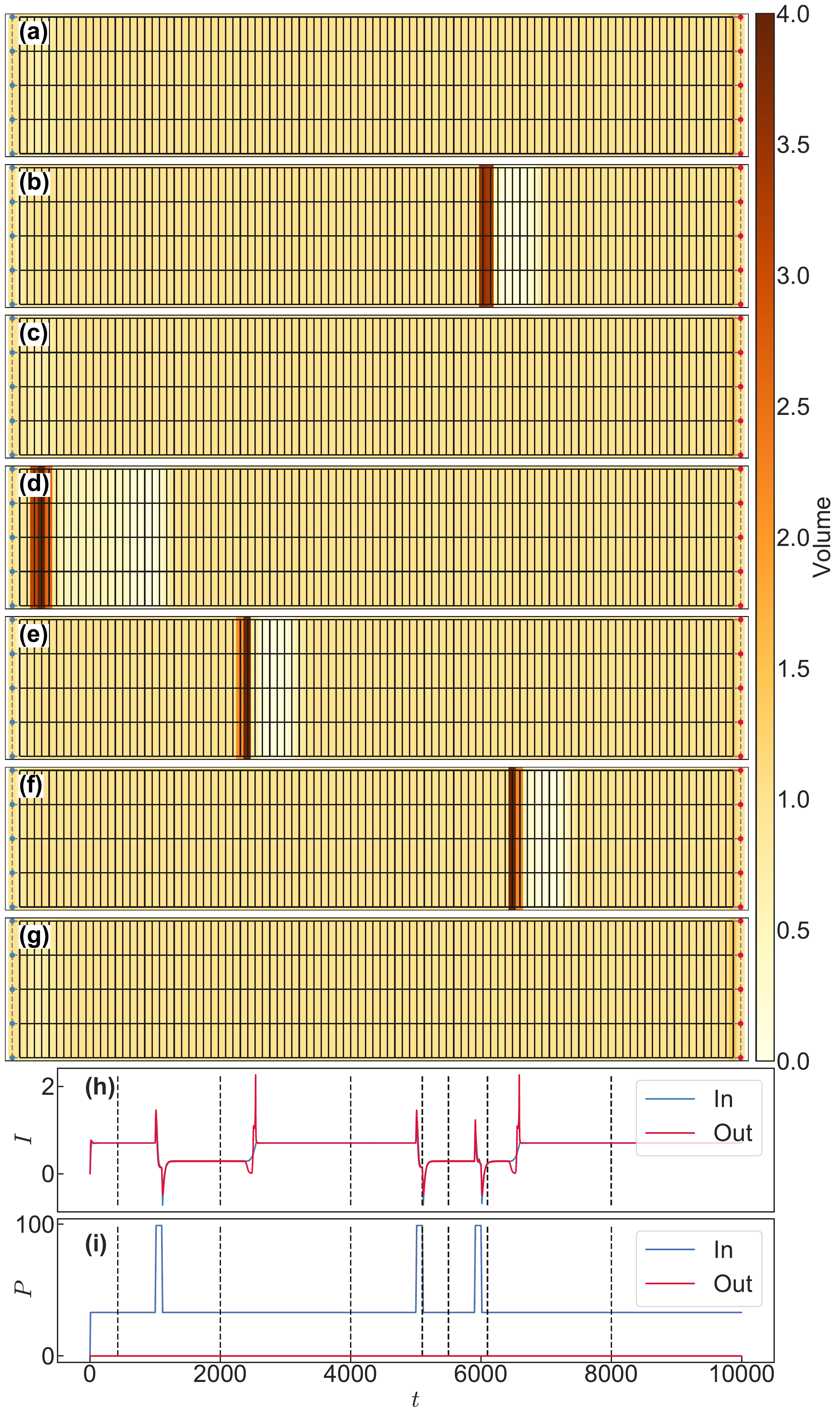}
    \caption{ Excitability of a network with $500$ nodes arranged in a rectangular grid, with $5$ rows and $100$ columns. The pressure of the red nodes in the right column (``Out'' nodes) is maintained constant at $0$ whereas the pressure of the nodes in the left column (``In'' nodes) follows the protocol displayed in blue in panel (i). Panels (a)-(g) display the volume at each node of the network for the different times marked in (h) and (i) with vertical dashed lines. Continuous lines in (a)-(g) stand for edges following $\Gamma_{NL}$ whereas dashed gray lines stand for $\Gamma_{L}$ (present only around the contacts).  $\epsilon=0.001$, $h=1/5$ and $\alpha = 1$. Panel (h) displays the current that goes in and out of the system as a function of time. For this configuration the system is stable around a state with a homogeneous volume distribution (panels (a), (c) and (g)). A short perturbation in the pressure (i) can trigger a pulse that propagates along the complete network. A second pulse cannot be excited while one is still travelling through the network (f).}
    \label{fig_excitability}
\end{figure}

In this work we have focused more extensively in the oscillatory regime of the system where time independent pressure boundary conditions result in time dependent behavior. However, one of the distinctive features of this model is its capacity to get exited by external perturbations. Excitable media, according to classical definitions, e.g. ~\cite{cross1993pattern}, show large excursions in phase space after being driven away from an equilibrium point, for a certain range of their parameters. In this subsection we show how our system responds to a pressure perturbation while within the excitable regime of the system.

In figure \ref{fig_excitability} we show a simulation in a rectangular network with $5 \times 100$ nodes. Panels (a)-(g) present snapshots of the volume distribution in the system at different times, whereas panel (h) shows the current that is going in and out of the system, and panel (i) presents the pressure at the contact points versus time. In the simulation, we rapidly increase the pressure at the blue nodes and keep it constant at $P=40$, a stable point of the system with an homogeneous volume distribution (Fig. \ref{fig_excitability} (a)). We then perturb the system with a brief increase of the pressure on the boundary. This triggers a pulse that travels through the system (a large excursion in phase space, see panel (b) of figure \ref{fig_excitability}). After the pulse arrives to the other end of the network (the low pressure contact points), the system is in its stable point again (panel (c)). We then trigger another pulse (Fig. \ref{fig_excitability} (d)), and while it is travelling through the system we introduce a third perturbation. However, the presence of the previous pulse prevents the creation of a new one and gives rise to an effective ``refractory'' time for the traveling excitation (see figure \ref{fig_excitability} (f)). Finally, with no more perturbations the system returns to the equilibrium behavior again, after the last pulse have exited the low pressure contact points (Fig. \ref{fig_excitability} (g)).

\subsubsection{The spatial footprints of travelling waves}

\begin{figure}
    \centering
\includegraphics[width=0.95 \linewidth]{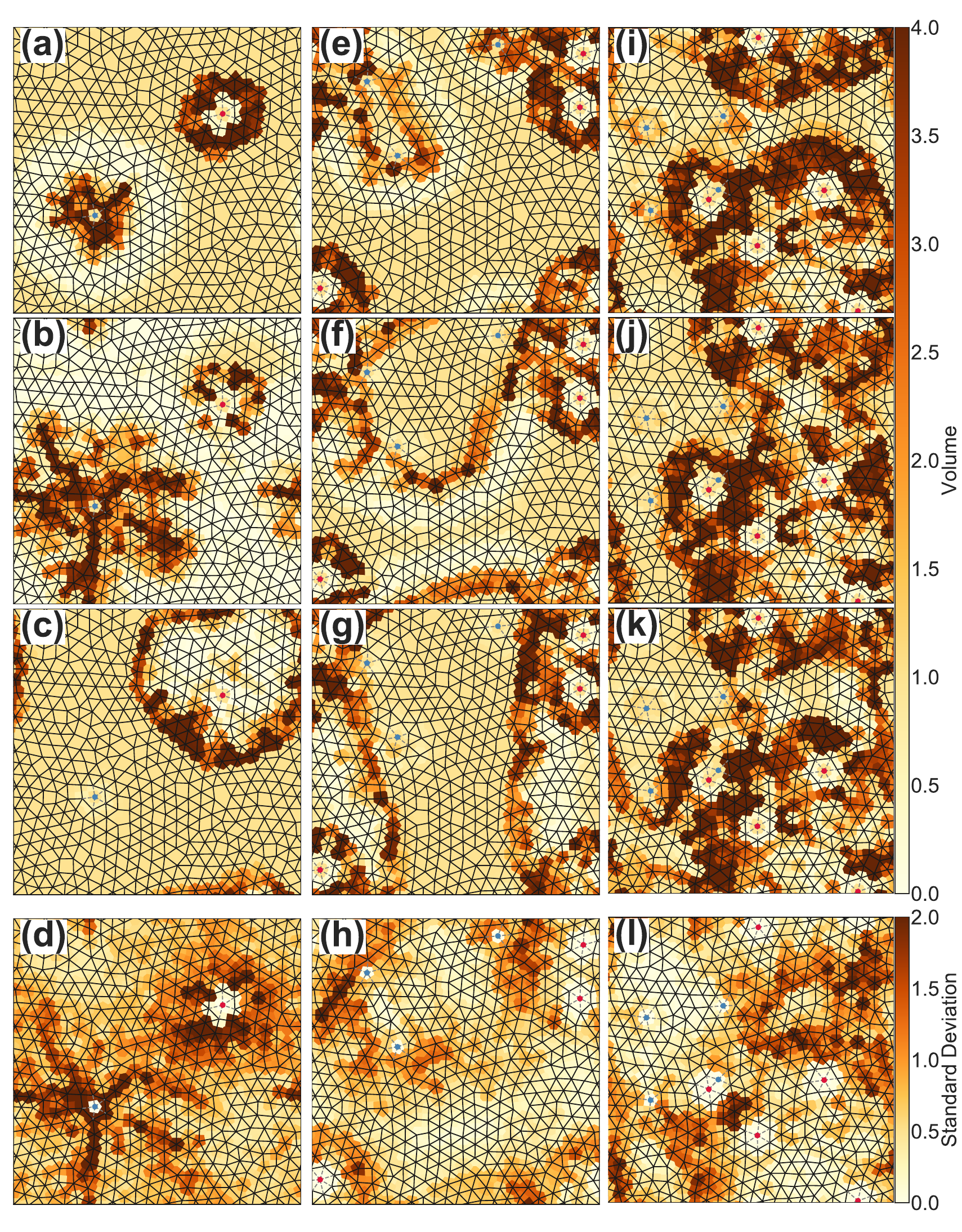}
    \caption{ Self-sustained oscillations for different sets of contact points. Rows 1 to 3 (panels a, b, c, e, f, g, i, j, k) correspond to snapshots whereas the bottom row (panels d, h, l) corresponds to the standard deviation of the volume at each node (it is calculated from the volume time series at every node). The color coding stands for volume accumulated at each node. We use a disordered planar network with $512$ nodes, average connectivity $\sim 5.5$, and periodic spatial boundary conditions. Continuous lines stand for edges following $\Gamma_{NL}$ whereas dashed gray lines stand for $\Gamma_{L}$ (here present only around the contacts). We used $\epsilon=0.001$, $h=1/5$, $\alpha = 0.32$ and $\Pi=150$ for all the simulations. Pressure is maintained constant at $150$ at the blue nodes, and $0$ at red nodes. Left column (panels a, b and c) presents waves travelling from one high pressure contact to a low pressure one. Central column (panels e, f and g) shows snapshots of traveling waves for $6$ randomly distributed contacts points ($3$ corresponding to high pressure and $3$ for low pressure). Right column (panels i, j and k) presents an analogous case but with $10$ contacts points ($5$ corresponding to high pressure and $5$ for low pressure). The bottom row (panels d, h, l) shows the profiles obtained after computing the standard deviation of the volume time series at each node. This is done after the initial transient has passed and the system displays stable oscillations.}
    \label{fig_averages}
\end{figure}

When waves travel through networks of non-linear resistors, they follow complex spatial-temporal patterns that depend on the network topology and position and number of contact points, something not present in the 1D analysis. In figure \ref{fig_averages} we show three different sets of pressure boundary conditions for a disordered planar network: $2$, $6$ and $10$ contacts (one case per column). Each configuration produces a different oscillatory pattern, where the volume stored in some nodes oscillates with a large amplitude, whereas the volume stored at other nodes is almost stationary. We display three snapshots for every configuration, panels (a-c) for the case with two contacts, panels (e-g) with $6$ contacts, and panels (i-k) with $10$ contacts. The bottom row (panels (d), (h) and (l)) displays the standard deviation of the time series of the accumulated volume at each node. In simple cases with a small number of contact points, these static profiles have spatial distributions that resemble the temporal-spatial patterns shown in the snapshots. In particular, conservation of mass imposes that pulses that change their shape increase their amplitude as they concentrate in smaller regions. This causes the standard deviations (panels (d), (h) and (l)) to highlight regions close to the contacts, with shapes that resemble the pulse fronts. For the case with two contacts (a-c), note that the volume wave front near the low pressure contact point is radially symmetric, but near the high pressure contact point the profile is dendritic. As the number of contact points increases then the oscillatory patterns become more complex. The standard deviation of the time-dependent volume stored at each node is highly variable. The magnitude of the fluctuations does not follow the simple patterns of the two contact case. Instead, we find regions close to the contact points that oscillate strongly, and regions close to them that are stationary. We hypothesize that these complex spatiotemporal patterns are partially due to constructive and destructive interference of the traveling waves, but the detailed study of the patterns is not in the scope of this work.

\subsubsection{Regions of linear conductance }

\begin{figure}
    \centering
\includegraphics[width=0.95\linewidth]{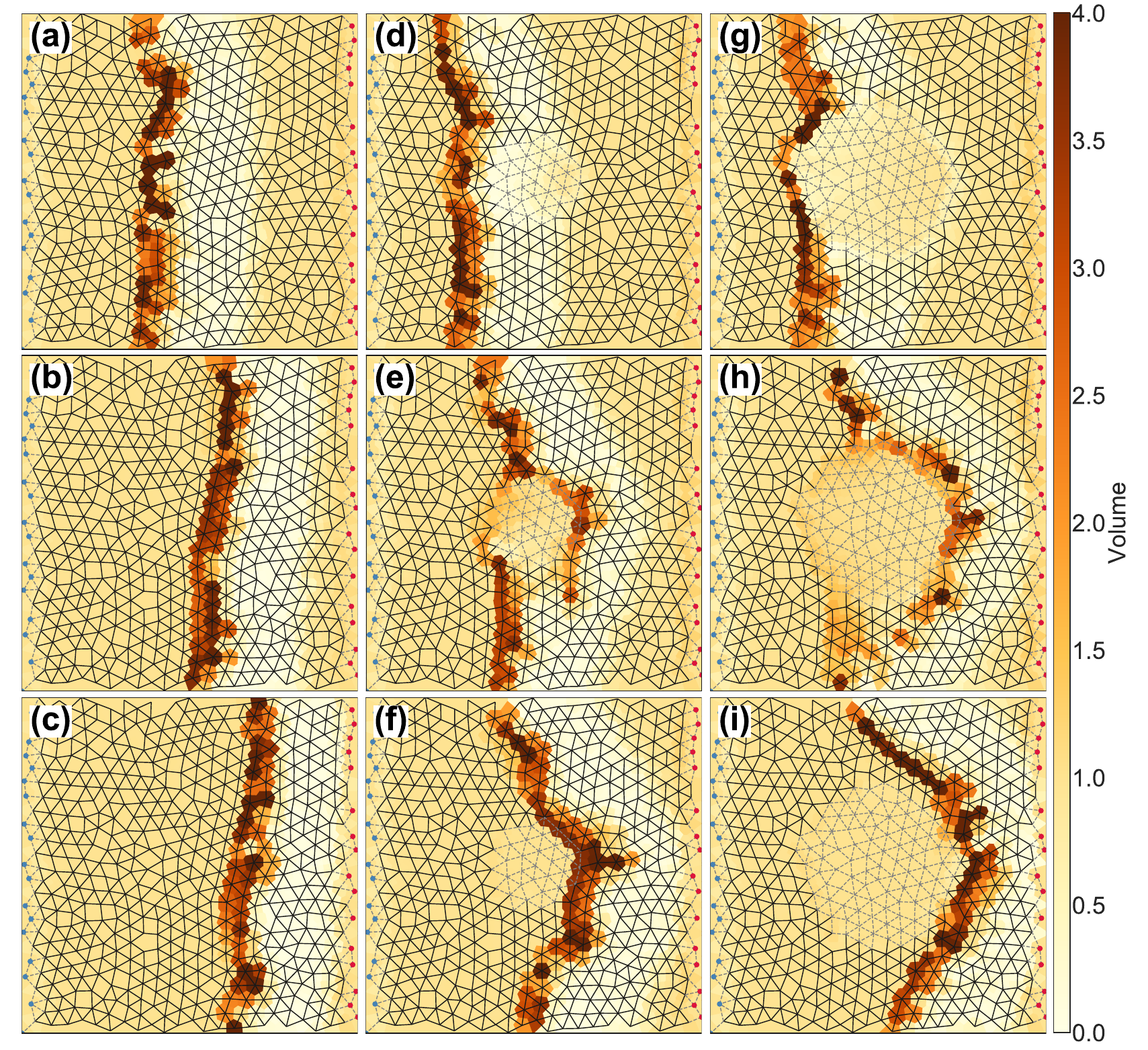}
\caption{Snapshots of planar waves traveling through a region of linear conductance. Color patches around each node stand for volume accumulated at that node. We use a disordered planar network with $512$ nodes and average connectivity $\sim 5.5$, without spatial periodic boundary conditions. Continuous lines stand for edges following $\Gamma_{NL}$ whereas dashed gray lines stand for $\Gamma_{L}$. We used $\epsilon=0.001$, $h=1/5$, $\alpha = 0.32$ and $\Pi=35$ for all the simulations. Pressure is maintained constant at $35$ at the left blue nodes, and $0$ at the right red nodes. (a)-(c) panels present snapshots of planar waves travelling from left to right. (d)-(f) panels show how the planar wave gets distorted when it reaches a central region of linear edges ($35$ linear edges). (g)-(i) present an analogous case but with $98$ linear edges in the central region. The waves ``leap frog'' the linear edges. This leads to an increase of the oscillation frequency, as the waves can traverse and exit the system faster.}
\label{fig_linear_central_edges}
\end{figure}

As we have discussed in other sections, it is the combination of the non-linear conductance and the coupling between volume and pressure that give rise to complex dynamics. In this section, motivated by \cite{Zykov2017}, we study how travelling waves interact with a region of linear edges. To achieve this we take a disordered planar network without periodic spatial boundary conditions (see Fig. \ref{fig_linear_central_edges}). We impose a constant high pressure to the contacts on the left boundary of the system and zero pressure to the contacts on the right. After a short transient, approximately flat fronts move from left to right, see snapshots in \ref{fig_linear_central_edges} (a)-(c). Now we modify the conductance of the edges in a circular region in the middle of the network, making them linear (following $\Gamma_L (\Delta P)$). When the fronts arrive to the linear region, they ``leap frog'' ahead and continue their propagation at the other side, see panels (d)-(f) of Fig. \ref{fig_linear_central_edges}. Finally, we do the same with a larger region in panels (g)-(i). These results are consistent with linear regions being areas of very fast pulse propagation. The observed behavior is reminiscent of that shown in Fig.~3A-C of Ref.~\cite{Zykov2017}, further strengthening the connection of the phenomenology of the model with standard excitable systems.

\section{Discussion}
\label{sec_discussion}

The work contained in this paper presents a model to study dynamics on complex networks. We use general phenomenological expressions that can be applied to a broad variety of problems. Indeed, these expressions can be modified and adapted to make them better approximate the governing equations of other physical or biological systems. We therefor expect the framework presented in this work to open new research avenues in the study of dynamics in non-linear flow networks of arbitrary topology. 
One such potential example is the spontaneous fluctuations of blood volume in the brain vasculature \cite{fox2007spontaneous}. It has been proposed that spontaneous fluctuations (in resting state) may be due to a non-neural origin~\cite{winder2017weak}, in contrast to typical brain hemodynamics which is driven by the activity of neurons. Understanding and modeling these phenomena in brain vasculature is of critical importance, since functional magnetic resonance imaging (fMRI) relies on the tight correlation of neural activity with blood volume and oxygenation. We have included in appendix \ref{sec_brain} a brief discussion of the physical arguments that may connect our model to brain hemodynamics. Brain blood flow dynamics is not the only biological system where spontaneous oscillations arise. Another system that involves intrinsic peristaltic-like contractions (which are also poorly understood) is the lymphatic system \cite{margaris2012modelling}.

In summary, this work shows how a network of nonlinear resistors can display emergent spontaneous dynamics for very different topologies and boundary conditions. The analytical results of section \ref{sec_math} help to understand the basic mechanisms behind the emergence of this complex behavior. We have shown how the negative-slope region makes the ``trivial'' homogeneous solution unstable in some cases, and how the system can support travelling waves. More detailed analysis of similar models in $1$D can be found in the semiconductor heterostructure literature, see e.g.~\cite{bonilla2005non,bonilla2010nonlinear}.

In \ref{sec_num_int} we discuss how to numerically integrate the system of equations for the case of a network of arbitrary topology. To do so we show how to include the pressure boundary conditions as a redefinition of the graph Laplacian. This enables us to integrate the system numerically in a straight forward way, obtaining the time-dependent pressure and volume at each node. This is a completely different approach than the one used to time integrate the equations of the models studying semiconductor superlattices~\cite{bonilla2005non}. Since those cases were $1$D the integration could be performed using one Lagrange multiplier. Our approach is more general and suitable for networks of arbitrary topology. 
In addition, our model opens the possibility of exploring other types of complex dynamics in flow networks, as it provides a general framework to explore systems with different expressions for edge conductance or for the volume-pressure coupling.

We have extensively discussed how the combination of non-linear edges and the coupling between pressure and volume can give rise to emergent spontaneous fluctuations under time-independent pressure boundary conditions. These systems present a broad array of interesting phenomena that will encourage further research, like the complex spatial patterns of volume fluctuations or the travelling wave behavior in inhomogeneous media composed of regions of linear edges. Moreover, in section \ref{sec_exc} we discussed how this model presents some properties which are typical of an excitable medium~\cite{cross1993pattern}, while still in the realm of distribution network theory. In this way we believe this model is an example of a new class of excitable systems, different from other models of excitable networks that explicitly use excitable elements in their nodes~\cite{kinouchi2006optimal,roxyn2004selfsustained}. Instead, our excitable flow network is composed of edges that present a nonlinear conductance and nodes that can store volume. The excitable nature of the system emerges as a product of the global coupling between currents, volumes and pressures.

\section{Acknowledgements}

This research was supported by the National Science Foundation via Award No. DMR1506625 (M.R.-G.), and the Simons Foundation via Award No. 454945 (M.R.-G.). E.K. acknowledges partial support by NSF Award PHY-1554887, the University of Pennsylvania Materials Research Science and Engineering Center (MRSEC) through Award DMR- 1720530, the University of Pennsylvania CEMB through Award CMMI-1548571, and the Simons Foundation through Award 568888.

\appendix

\section{Coupling between volume accumulation and pressure in flexible tubes} 
\label{sec_coupling_pres_vol}

\begin{figure}
    \centering
\includegraphics[width=\linewidth]{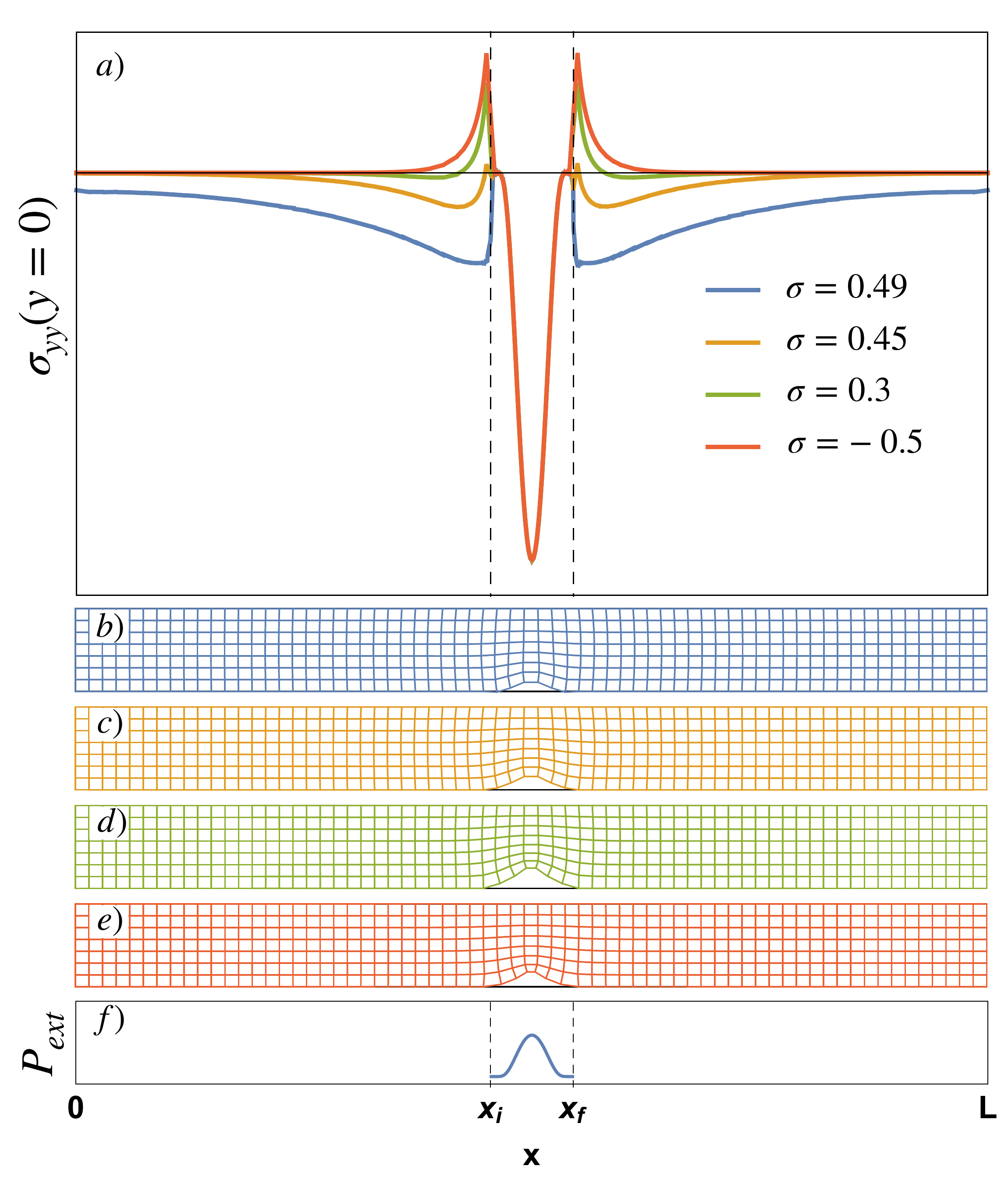}
    \caption{Idealized deformation of the medium surrounding a volume accumulation inside a flexible tube and associated stresses. We consider rectangular domains with different Poisson ratios. The boundaries are clamped (displacements, $u_1$ and $u_2$, are set to zero), except for a free ``surface'' ($y=0$ and $x_i < x< x_f$) where we impose an external pressure (panel f). In panel (a) we measure the pressure that the medium imposes on the vessel, $\sigma_{yy}(y=0)$. All lines collapse in the free surface region, canceling the imposed external pressure. Outside this region different Poisson ratios produce different pressure profiles that decay away from the free surface region. For large Poisson ratios the deformation of the surrounding medium produces a pressure field of negative sign (against the tube), qualitatively analogous to the pressure-volume relation that we use in our model. The displacements present in each of the cases are displayed in panels (b)-(e), where the meshes have been simplified for easier visualization. Dashed lines in (a) and (f) mark the region where the external pressure is applied.}
    \label{fig_deformations}
\end{figure}

We consider a network of flexible tubes  embedded in an elastic (almost incompressible) medium. The deformation of the tube wall and the surrounding medium controls the pressure response to a local volume accumulation. This relation is included in our model through the phenomenological relation \eqref{eq_vol_pres}. This equation is non-local, which means that an accumulation of volume causes a increase of pressure not only in the region of accumulation but also in neighboring sites. In particular, a local volume accumulation in relation \eqref{eq_vol_pres} produces a pressure field that decays with distance from the region of accumulation.

Here, we test whether a local volume accumulation inside a hollow vessel embedded in an elastic medium could produce a decaying pressure field. In particular, we solve the equilibrium equations of classical linear elasticity~\cite{landau1986course},
\begin{equation}
    \frac{E}{2(1+\sigma)} \frac{\partial^2 u_i}{\partial x_k ^2} + \frac{E}{2(1+\sigma)(1-2\sigma)} \frac{\partial^2 u_l}{\partial x_i \partial x_l } = 0,
\end{equation}
where $u_i$ are the components of the displacement vector field, $E$ is the Young's modulus, $\sigma$ is the Poisson's ratio, and $x_i$ are the spatial variables; summation over repeated indices is implicit. 

For simplicity, we use a rectangular domain as a $2$D version of our problem, and we clamp all its boundaries except for the region $ \partial \Omega _{free} \equiv (y=0, \ x_i<x<x_f$), see figure \ref{fig_deformations}. The lower boundary ($y=0$) of the elastic medium is meant to represent the interface between the tube (that carries the fluid) and the medium that embeds it. We impose a vertical pressure on the free ``surface'' ($\partial \Omega _{free}$) with a bump shape, $P_{ext} = 30  e^{(-1/(0.25 - ((x - x_i)/(x_f - x_i) - 0.5)^2))}$, displayed on figure \ref{fig_deformations}~(f). As the rest of the lower boundary is clamped, we can measure the pressure that the medium exerts on the non-deformed region of the tube. In the presence of fluid in the tube, this pressure field would promote flows that would give rise to new deformations, a phenomenology that is described in the rest of this work.

Finally, the pressure field exerted by the medium on its lower boundary ($\sigma_{yy}(y=0)$) is displayed in figure \ref{fig_deformations}~(a). To illustrate this effect, we use four different values of the Poisson's ratio, although only values close to $0.5$ are probably relevant in most experimental cases. All lines collapse in $\partial \Omega_{free}$, as expected since they need to cancel the externally applied pressure. Outside this region ($x<x_i, \ x_f<x$), there is a pressure field whose magnitude decays with the distance to the volume accumulation, and the direction depends on the sign of $\sigma$. For positive Poisson ratios (the most relevant situation) the direction of the pressure response is against the tube wall, in qualitative agreement with expression \ref{eq_vol_pres}. For visualization, we also plot the deformations undergone by each material using a simplified mesh, Fig. \ref{fig_deformations}~(b)-(e).

\section{Possible sources of non-linearities in biological systems}

\label{sec_brain}

The model contained in this work describes the emergence of complex dynamics in flow networks, allowing for local volume accumulation within the system and non-linear conductances for the edges. In this appendix we describe various phenomena in the mammalian brain vasculature that could potentially produce complex non-linear behavior for the conductance of the vessels, reminiscent of the non-linearities present in $\Gamma (\Delta P)$.

Different vessels present in the mammalian brain vasculature display a broad and complex response to changes in pressure or flow conditions. This response can be active, when the vessel modifies its muscle tone, or passive, controlled only by the fluid-mechanical interaction between the blood and the vessel. 
A detailed account of all effects lays outside the scope of this work. However here we present a short review for the interested reader.

\textit{Active nature of vessels}. Since the seminal work of Bayliss in 1902 \cite{bayliss1902local}, it is known that vessels can present a myogenic response, as they constrict in response to an increment of internal pressure. Flow has also been experimentally found to cause dilation and contraction of vessels \cite{bevan1991pressure}. This response depends on different factors, such as the internal pressure \cite{thorin_trescases1998high} and the ability of endothelial cells to sense blood flow \cite{yamamoto2006impaired}. Experimental work \cite{thorin_trescases1998high,ngai1995modulation} is consistent with a non-monotonic $\Gamma(\Delta P)$ function. The dilation of the vessel in response to sheer stress is non-monotonic for low an intermediate myogenic tone. As sheer stress increases, the vessel dilates until it reaches a maximum radius and then reduces the radius for larger sheer stress. That causes a non-monotonic relation between flow and pressure difference.
In addition, some experimental work has shown an oscillatory myogenic response to a constant internal pressure~\cite{osol988spontaneous}. We do not consider this effect in our model although it could be included as edges presenting a time-dependent $\Gamma(\Delta P)$.

\textit{Passive response}. 
It has been theoretically proposed  \cite{kumaran1995stability} that a viscous flow through a flexible tube can become unstable. When the pressure difference between the ends of the tube reaches a critical value, any small perturbation in the flow will exponentially grow producing a deformation of the flexible tube and making the fluid flow depart from the laminar behavior. This sudden change increases the energy dissipated in the system and results in a consequent drop of the total flow.
This type of instability has been experimentally measured in \cite{kumaran2000spontaneous,neelamegam2014instability}, where the authors observed a sudden increase of the effective viscosity of the fluid due to the development of the instability. This would be consistent with the non-monotonic flow-pressure relation used in this work. The critical velocities for which the linear stability analysis of \cite{kumaran1995stability} gives the first unstable mode is of the order of cm/s for a vessel with a diameter of $100$ $\mu$m and of the order of mm/s for a $5$ $\mu$m capillary, in good agreement with typical blood velocities, \cite{secomb2017blood}.  Nonetheless, this passive response still needs to be measured experimentally in real blood vessels. In addition to this, blood rheology may also play an important role since blood is a complex fluid which constituent agents are deformable and very often of the order of the vessel radius \cite{secomb2017blood,fung2013}.

\bibliography{bibliography_pre.bib}

\end{document}